\RequirePackage{fix-cm}
\documentclass[smallextended]{svjour3_tam}       
\smartqed  
\usepackage{graphicx}
\usepackage[square,sort,comma,numbers]{natbib}
\usepackage{color}
\usepackage{gensymb}
\usepackage{parskip}
\usepackage{hyperref}
\newcommand{\bi}{\begin{itemize}}
\newcommand{\ei}{\end{itemize}}
\newcommand{\beq}{\begin{equation}}
\newcommand{\eeq}{\end{equation}}
\newcommand{\bea}{\begin{eqnarray}}
\newcommand{\eea}{\end{eqnarray}}
\newcommand{\bqu}{\begin{quote}}
\newcommand{\equ}{\end{quote}}
\newcommand{\bctr}{\begin{center}}
\newcommand{\ectr}{\end{center}}
\newcommand{\bd}{\begin{description}}
\newcommand{\ed}{\end{description}}
\newcommand{\bdm}{\begin{displaymath}}
\newcommand{\edm}{\end{displaymath}}
\newcommand{\lsim}{\mbox{$\:\stackrel{<}{_{\sim}}\:$} }
\newcommand{\gsim}{\mbox{$\:\stackrel{>}{_{\sim}}\:$} }

\newcommand{\orr}{\rm \Omega_r}
\newcommand{\om}{\rm \Omega_{\rm m}}

\newcommand{\oll}{\rm \Omega_\Lambda}
\newcommand{\omol}{\rm \Omega_{\rm m},\Omega_\Lambda}
\newcommand{\ob}{\rm \Omega_{\rm b}}
\newcommand{\oc}{\rm \Omega_{\rm c}}
\newcommand{\obhh}{{\rm \Omega_{\rm b}} h^2}
\newcommand{\ochh}{{\rm \Omega_{\rm c}} h^2}

\newcommand{\zp}{\ensuremath{z_{\rm p}}}
\newcommand{\zo}{\ensuremath{z_{\rm o}}}

\newcommand{\zbar}{\ensuremath{\bar{z}}}

\newcommand{\kmsmpc}{\ensuremath{{\rm km\, s}^{-1}{\rm Mpc}^{-1}}}

\newcommand{\blue}{\color{blue}}

\def \aap{A\&A}

\def \aj{AJ}

\def \apj{ApJ}
\def \apjl{ApJ}
\def \apjl{ApJ Lett.}
\def \apjs{ApJS}

\def \azh{Astronomicheskii Zhurnal}

\def \jcap{J. Cosmo. Astropart. Phys.}

\def \mnras{MNRAS}

\def \nat{Nature}

\def \physrep{Physics Reports}

\def \prd{Phys. Rev. D}

\def \sovast{Soviet Astronomy}

\begin{document}

\title{Cosmological constraints on dark energy\thanks{TMD acknowledges the support of the Australian Research Council through Future Fellowship, FT100100595, \& the ARC Centre of Excellence for All Sky Astrophysics, CE110001020.}
}


\author{Tamara M.\ Davis 
}


\institute{T.~M.~Davis \at
              School of Mathematics and Physics, University of Queensland, QLD 4072, Australia \\
              ARC Centre of Excellence in All Sky Astrophysics (CAASTRO) \\
              \email{tamarad@physics.uq.edu.au}           
}

\date{Received: 24 January 2014 / Accepted: 23 March 2014}

\maketitle

\begin{abstract}
It has been only $\sim$15 years since the discovery of dark energy (although some may argue there were strong indications even earlier).  In the short time since measurements of type Ia supernovae indicated an accelerating universe, many other techniques have now confirmed the acceleration is real.  The variety of ways in which dark energy has been confirmed is one of the reasons we are so confident in the statement that most of the energy in the universe is in a form we can not see except through its gravitational influence.  

This review aims to summarise briefly the many varied ways we now have measured dark energy.  The fact that these different techniques all indicate that the simplest model remains the best -- that dark energy contributes a constant background acceleration -- is remarkable, since each of these different types of measurements represented opportunities for this simplest model to fail.  Although we currently lack a compelling theoretical explanation for this acceleration, any explanation will have to explain the wide variety of complementary observations that we review here.  

This is an informal presentation, following the lines of the talk I presented at the General Relativity and Gravitation (GR20) conference in Warsaw in July 2013.

{\blue This astro-ph version contains bonus material, indicated by blue text, that would not fit within the published version's page limits.}
\end{abstract}

\newpage
\section{Introduction}
\label{intro}
\vspace{-5mm}
We're at an exciting point in the link between cosmological observations and fundamental physics.  We've discovered that the universe is accelerating.  The cause of that acceleration we give the name {\em dark energy}.  However, we don't know what that dark energy actually is.  Could it require alterations to the laws of gravity?  Is it a fluid with negative pressure?  And if it is vacuum energy, could a quantum theory of gravity naturally explain why its value is $\sim$ 54 orders of magnitude larger than the naive value quantum physics predicts?\footnote{The fabled 120 orders of magnitude that is usually quoted is apparently an exaggeration --- see \cite{martin12}, around Eq.~548.  Nevertheless, this remains one of the worst ever matches between theory and observation.} 

Since the discovery in 1998 of the acceleration of the expansion of the universe \citep{riess98,perlmutter99}, many new and varied measurements have confirmed its existence.  And despite many chances to find otherwise, every observation so far has confirmed that it is consistent with a cosmological constant -- uniform throughout space, it doesn't clump and does not dilute as the universe expands.  This strange behaviour means that it could be a form of vacuum energy, of the type that is predicted to exist by quantum physics.  The only problem is that quantum physics predicts a larger amount of this vacuum energy than we actually see.  Whether or not that is a problem may depend on whether the successful merging of quantum physics and gravity into a quantum theory of gravity can explain away that conundrum. 

In any case, a decade and a half of observations have provided numerous new {\em types} of measurements, which could potentially have revealed discrepancies from the vacuum energy / cosmological constant picture of dark energy.  However, in every case, the observations have remained consistent with vacuum energy -- the simplest form of dark energy possible.   The next generation of experiments could potentially reveal the same thing.  If so, we will be none the wiser.  The real breakthrough in dark energy research at this point has to be theoretical.  Even if observers do find some discrepancy with vacuum energy, it may still be that a compelling theoretical explanation is lacking.  

However, it is very difficult to make a theory of dark energy that differs from a cosmological constant, without violating one observation or another.  The purpose of this review, is to summarise the many varied observations that now confirm dark energy, with a view to guiding theorists in the many ways that their theory has to pass observational tests if it is to be successful.  It is through this communication between observation and theory that we will be able to advance our knowledge of this enigmatic feature of our universe. 

This is an informal review following the lines of the talk I presented at the General Relativity and Gravitation (GR20) conference in Warsaw in July 2013.  I aim to fill the gap between technical papers and popular accounts.  As such this is a primarily qualitative review.  For more technical details I highly recommend the recent review \cite{weinberg13}.

\vspace{-3mm}
\section{Background: The standard $\Lambda$CDM model}
\vspace{-5mm}
Throughout I assume some familiarity with the basic terminology of cosmology, such as redshift $z$, but give a brief introduction here.  The standard cosmological model, $\Lambda$CDM, consists of approximately 30\% matter and 70\% cosmological constant, $\Lambda$.  Most of the matter is cold (non-relativistic) and dark (does not interact electromagnetically), so is known as cold dark matter (CDM).  It is expanding at about $H_0=70\kmsmpc$ where Hubble's constant $H_0$ is the ratio of the expansion rate $da/dt$ to the scalefactor $a$ (relative size of the universe normalised to 1 at the present day).  Redshift and scalefactor are related by $1+z=1/a$.  Often it is useful to express measurements, in a way that is independent of the uncertain value of $H_0$, so we use $h=H_0/100\kmsmpc$ and express distances in Mpc$/h$.  

Both the expansion rate and scalefactor change with time, so Hubble's `constant' is redshift dependent $H(z)$.  How the expansion rate changes is determined by the gravitational effect of the components of the universe, such as the matter density, cosmological constant density, and radiation density, denoted $\om, \oll$, and $\orr$ when normalised to the critical density at the present day (the density required to make the universe spatially flat).  Observations show that the universe is close to flat so $\om+\oll+\orr\sim1$.    

Common variations on $\Lambda$CDM include $w$CDM, in which the equation of state (the ratio of pressure to energy density) of dark energy $w=p/(\rho c^2)$ is allowed to take a value different from $w=-1$; $w(a)$CDM in which $w$ can change as the universe expands; and the flat versions of all the above models, where $\sum{\rm \Omega}_i\equiv1$ is assumed. 

Note that I use the term {\em dark energy} to refer generically to whatever explains the observations of the accelerating universe.  That could be a modification to the theory of general relativity (e.g.\ $f(R)$ gravity, galileons), a new fluid that reacts repulsively to gravity (e.g.\ vacuum energy, quintessence), or the possibility that our assumption of large-scale homogeneity and isotropy has led us to false conclusions (such as in inhomogeneous cosmological models).   A cosmological constant and vacuum energy are equivalent, and represent the simplest model of dark energy, in which dark energy provides a uniform acceleration that doesn't dilute as the universe expands.\footnote{They are equivalent in their effect on the expansion, but differ only in which side of the general relativistic field equations they appear; the cosmological constant as part of the theory alongside $G_{\mu\nu}$, and vacuum energy as a component of the energy density in $T_{\mu\nu}$.}

The strength of cosmological constraints comes from the complementarity of the many different types of observations described below.  In particular, the cosmic microwave background (CMB) is very important, and most of the other observables give strong constraints only when combined with some information from the CMB.  Equally, the CMB alone can not put strong constraints on all parameters.  For example, it gives weak constraints on dark energy and Hubble's parameter (see lowest panel of Fig.~21 in \cite{planck13_XVI}).  Dark energy dominates in the low-redshift (late time) universe, but matter or radiation dominated in the high redshift (early time) universe.  So it is by combining low-redshift constraints like large-scale structure and supernovae with high-redshift constraints like the CMB, that our full model of the universe is best formed. 

\vspace{-5mm}
\section{Fertile ground}
\vspace{-3mm}

It may seem like the discovery of dark energy popped into existence in 1998, with the announcement that distant supernovae appeared fainter than they should if the universe was decelerating \citep{riess98,perlmutter99}.  Certainly the media enjoy portraying it as a shock and, when waxing lyrical, researchers (including myself)  often say that dark energy goes against everything we thought we knew about how gravity should work.  Nevertheless, the idea did land on fertile ground.  

The cosmological constant was, of course, introduced by Einstein to allow for a static universe \citep{desitter17,einstein17}, after it was pointed out that his new theory of general relativity meant that the universe was unstable \citep{desitter17,friedmann22,lemaitre27,eddington30}.  It had to be expanding or contracting, which went counter to the compelling evidence of the time that the stars were not in general moving towards or away from us.  Although Slipher \citep{slipher15} had actually published data in 1915 showing that the majority of `spiral nebulae' were redshifted, and therefore had velocities away from us, the discovery of the linear velocity-distance relation (Hubble's law) had to wait for catalogues of distance measurements made using the period-luminosity relation of cepheid variables \citep{leavitt12} 
 before Hubble could publish his seminal paper in 1929 \citep{hubble29}.   

The cosmological constant was then considered defunct, but came in and out of vogue several times as subsequent observations suggested it was required, and then were discredited.  

That's probably why the evidence in the early 1990's that a cosmological constant might exist, was treated with extreme skepticism.  However, there was already fairly strong evidence that an accelerating universe may have been necessary to explain the existing data \citep{efstathiou90,ostriker95,krauss95,yoshii95}.  

Some of the most straightforward tests were {\bf galaxy number counts}.  Galaxies were too numerous at high-redshift.  The number in a particular volume of space in the distant universe was greater than the number in the same size volume of space in the local universe.  Some observers were adamant that galaxy merging could not be rapid enough to get rid of that many galaxies, and if the numerator was correct, perhaps the denominator was flawed \citep{yoshii95}.  They suggested that there must actually be more volume out there than our naive calculations indicated, which would be the case if the universe were accelerating.\footnote{Yoshii \& Peterson \citep{yoshii95} say in their abstract ``we find that these number count data favor a flat, low-density ${\rm \Omega}_0\sim0.2$ universe with a nonzero cosmological constant''.  Thus presenting evidence for a non-zero cosmological constant, of approximately the now-accepted magnitude, three years before the supernova results.}  By assuming a decelerating universe we could have incorrectly calculated distances to these galaxies, and thus underestimated the volume across which they were spread.  However, that was too hard a nut to swallow.  It was generally believed that it was our incomplete knowledge of galaxy evolution, not any weakness in our theory of gravity, that caused the overabundance of distant galaxies \citep{campos96}.  Other galaxy measurements, even earlier, \cite{efstathiou90} had presented evidence that showed galaxy clustering was too strong on large scales, and argued for ``a spatially flat cosmology in which as much as 80\% of the critical density is provided by a positive cosmological constant'' --- prescient words 8 years before the supernova studies came to approximately the same conclusion.

Next there's the {\bf missing density} problem. When we cast our mind back to the state of cosmology in the early 1990's, you realise just how far we have come in those two short decades.  Nowhere is this more starkly evident than in a bet made between PhD officemates Sean Carroll and Brian Schmidt in 1991.  They bet that the density of the universe, would not be known to better than $\pm0.3$ by 2011.  
By 2011 the error bars we were quoting on the matter density of the universe were actually on the order of $\pm0.017$, i.e.\ 6\% (for $\ob+\oc$ for {\em WMAP}+BAO+$H_0$ in Table 1 of \cite{komatsu11}).  The reason that this accuracy would be such a surprise to someone from 1991, was that back then there remained a strong dispute over whether the universe was significantly underdense or flat.

Galaxy observers had measured the mass in galaxies (including dark matter) and come up short.  There was not enough matter in the universe to make it flat.  In fact they could only find about 30\% of the critical density of matter, ${\rm \Omega}\sim0.3$.  Meanwhile, theorists enthusiastic about inflation, were convinced that the universe had to be very close to flat indeed, for anything else was unstable (not an attractor).  In a manner rather reminiscent of ``I would have had to to pity our dear God.  The theory is correct anyway.''\footnote{Isle Rosenthal-Schneider reported that this was Einstein's reply when asked what he would have done if the eclipse measurements had not confirmed his theory of general relativity \citep{rosenthal80}.}, they were themselves adamant that the universe should have the critical density, ${\rm \Omega}=1.0$.   Indications that the CMB, measurements of $H_0$, and nucleosynthesis constraints required a cosmological constant were evident as early as 1995 \citep{ostriker95}.  Later observations would also weigh-in on the ${\rm \Omega}\sim1.0$ side, as measurements of the cosmic microwave background would soon indicate that the universe was close to flat \citep{debernardis00,melchiorri00,hanany00}.

\begin{figure}\sidecaption
\resizebox{0.7\hsize}{!}{\includegraphics[width=84mm]{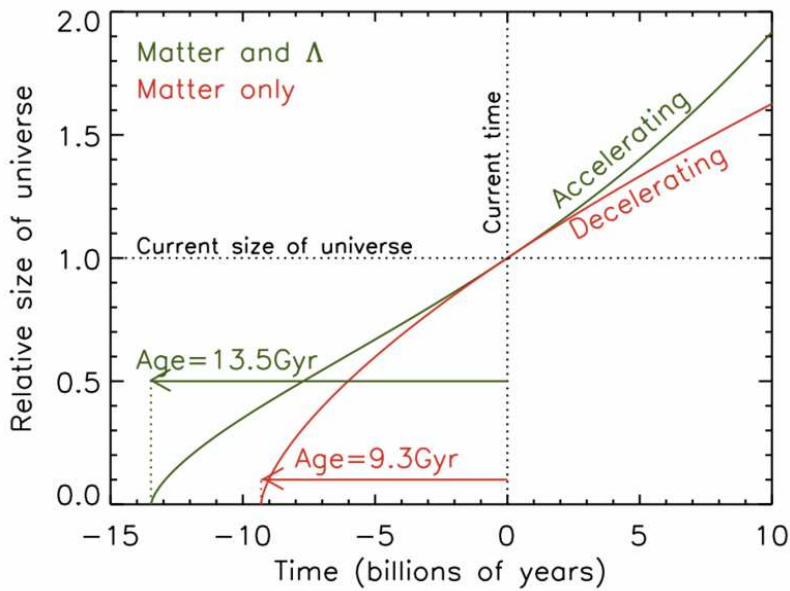}}
\caption{Relative size of the universe over time for two flat universes, one with matter density $\om=1.0$ and no dark energy (red) and the other with $\om=0.3$ and a cosmological constant $\oll=0.7$ (green).  If the universe decelerated to its current rate of expansion (red), it would only be about 9.3Gyr old, whereas if you add a period of acceleration then the universe is older, about 13.5Gyr for this model.   Both curves have been normalised to the current rate of expansion of the universe, assumed to be 70\kmsmpc.  \vspace{15mm}}
\label{fig:age}
\end{figure}

The breach of ${\rm \Omega}\sim0.7$ is why Mr Schmidt and Mr Carroll could consider placing a bet that ${\rm \Omega}$ would not be known to within $\pm0.3$ within 20 years.  They were betting that this disagreement between observation and theory would not be resolved.  

Brian's punishment for helping to discover dark energy was losing that bet (but winning a Nobel prize was probably a good exchange).  The discovery revealed that ${\rm \Omega}$ was not made up of matter density alone.  It also contained dark energy, which, if in the form of a cosmological constant or vacuum energy, contributed $\oll\sim0.7$ to the mix.  So ${\rm \Omega}$ acquired subscripts, and the main components of the universe were $\om\sim0.3$ and $\oll\sim0.7$.  The total density of the universe is what theorists really cared about, and so the fact that the dark energy almost perfectly filled the $0.7$ gap made everyone happy.

And then there's the big one.  Until 1998 it appeared that the universe was {\bf younger than the oldest stars}.  If the universe had decelerated to its current rate of expansion, then it would be about 9.3 Gyr old (flat universe with no dark energy), but we know that the oldest stars are over 13 Gyr old -- a problem in anyone's philosophy.  Perhaps we had the current rate of expansion wrong (measuring Hubble's constant, after all, does have a shaky history), or perhaps we didn't understand stars (even though, for better and worse, we humans are pretty good at calculating nuclear reaction rates).  Again, dark energy saves the day because if the universe accelerated to its current rate of expansion, then it would be older than if it had decelerated to its current expansion rate (see Fig.~\ref{fig:age}).  So the discovery of acceleration greatly satisfied those who were concerned that the universe should be older than the stars it contains.
 
So the number counts of galaxies, the predictions of inflation along with the measurements of the CMB, and the ages of the oldest stars, were the fertile ground into which the discovery of the acceleration of the expansion of the universe fell.  It is to the subsequent confirmations of that acceleration that we will now turn.

\vspace{-5mm}
\section{Supernovae}
\vspace{-3mm}

\subsection{Supernovae --- Measuring expansion rate}
\vspace{-5mm}

Supernovae have provided probably the most well known evidence for the acceleration of the expansion of the universe.  They can be calibrated to be standard candles \citep{phillips93,phillips99} -- sources of known luminosity -- which means that we can determine their distance from us by measuring their apparent brightness.  Couple that with the (much easier) measurement of their redshift, and you know how fast the universe was expanding back when the supernova emitted the light we're now seeing.  (Or if you prefer, how much the universe has expanded between then and now.)  Thus measuring the redshift of many supernovae at a range of distances allows us to measure how the rate of expansion of the universe changes over time.  Different amounts of matter and dark energy decelerate or accelerate the universe at different rates.  So by observing the magnitudes and redshifts of supernovae we can measure the amount of dark energy and matter. 

\begin{figure}\sidecaption
\resizebox{0.75\hsize}{!}{\includegraphics[width=84mm]{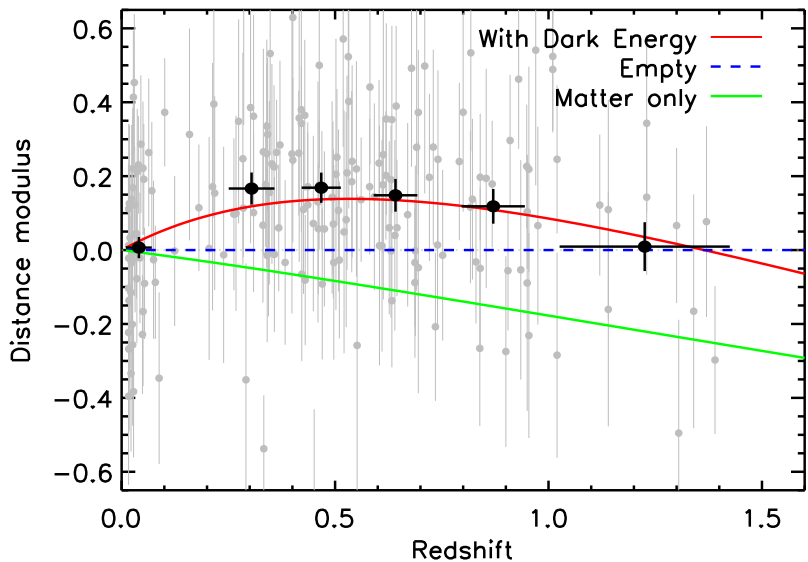}}
\caption{Hubble diagram --- supernova magnitude vs redshift --- normalised to the predicted expansion history for an empty universe (dashed line).  Raw data are in grey, binned with $1\sigma$ uncertainties in black.  The red line is a flat model with 30\% matter, 70\% dark energy.  The green line that is a poor fit to the data has 30\% of the critical density of matter and zero dark energy. (Adapted from Fig.~7 of \cite{davis07} --- there is now far more data than shown here.)
\vspace{12mm}}
\label{fig:sn}
\end{figure}

In practise, we plot the observed luminosity of the supernovae against their observed redshifts, and compare that to  the expected luminosity and redshift for different cosmological models (see Fig.~\ref{fig:sn}).
The evidence that the expansion of the universe is accelerating is really an observation that high-redshift supernovae are fainter than would be expected in a decelerating universe.  Initially it was questioned whether that faintness should be attributed to acceleration or some other phenomenon.  For example, some sort of intergalactic dust might be absorbing the radiation en route to our telescopes, falsely dimming the supernovae.  
To fit the dimming of supernovae, the proposed {\bf grey dust} must have no colour preference, and absorb all wavelengths equally.  This would be strange behaviour, considering the normal dust we find on Earth and in our Solar System  tends to redden a spectrum by preferentially absorbing or scattering bluer wavelengths.  Although such grey dust seemed unlikely, it was arguably less unexpected than a new law of gravity or negative-pressure fluid, and so was initially taken as a legitimate possibility.

Grey dust was quickly ruled out by observations of extremely distant ($z\gsim1$) supernovae that were brighter than expected.\footnote{Brighter than expected if pure acceleration or dust were the cause, but still fainter than a pure decelerating universe.  The first examples were SN 1997ff at $z\sim1.7$ published in 2001 by \cite{riess01} and 16 more at $z>1.25$ published in 2004 by \cite{riess04}.}  The $\Lambda$CDM model is one that was initially decelerating.  When one can measure far enough away one can see well into the decelerating regime, in which the supernovae should once again be brighter.\footnote{\blue Note that the transition from deceleration to acceleration, and the transition from matter to dark-energy dominance, do not occur at the same redshift.  In a flat $\Lambda$CDM universe with $(\omol)=(0.3,0.7)$ acceleration begins at $z=0.67$, while dark energy doesn't dominate the energy density of the universe until $z=0.33$.}  
 In contrast, dust would continue to make more distant supernovae ever dimmer (see Fig.~7 of \cite{riess04}).

The other main possibility that was floated was that we were living in a {\bf enormous underdensity}, sometimes referred to as a {\em Hubble bubble}.  The models of the universe that we use assume homogeneity and isotropy.  If this is violated, then the calculations of luminosity vs redshift would have been flawed.  In principle, an underdensity of appropriately tuned size and profile, can mimic almost any smooth magnitude-redshift distribution, so is very difficult to disprove with supernova data alone, but this had several philosophical problems.  Firstly for the model to be viable the underdensity had to be on the order of half of the size of the observable universe.  Secondly, we had to be pretty close to the centre of it, to avoid a strong dipole measurement in the CMB.  Being in the centre of such a large overdensity violated the Copernican principle, which states that we should not expect to find ourselves in a special region of the universe.  However, history has shown that the universe is not limited by our philosophical preferences,  so we prefer to base our conclusions on observations.  Other types of observations now strongly disfavour the large local void hypothesis, and support the presence of dark energy (see discussion in \cite{moss11,valkenburg13}).  


\vspace{5mm}
{\blue
{\bf Observational considerations}

The largest uncertainty in supernova studies arises in figuring out the magnitudes that should go on the magnitude-redshift diagram.  Type Ia supernovae brighten and fade over about a month.  The magnitude used as a standard candle is related to the ``peak magnitude'' of this light curve, when the supernova hits its brightest point.  The peak magnitude of type Ia supernovae has an intrinsic dispersion of about an order of magnitude, but this dispersion can be reduced using the width-luminosity relation --- the supernovae that take the longest to brighten and fade are also intrinsically brightest.  Using that information allows us to compute a calibrated standard candle with dispersion about 0.15 magnitudes.  

\vspace{5mm}
Measuring the peak magnitude is subtle in several ways.  Firstly, one rarely has an observation precisely at the moment of peak, and therefore one needs to fit a model light curve to the data points one does have.  Secondly, the peak is not defined as when the total flux is maximal, but when the flux in a certain passband is maximal.  Usually that is the magnitude in the $B$-band (a filter that transmits the blue end of the visible spectrum).  As supernovae are redshifted, different regions of their spectrum would land in the $B$-band, and therefore other filters at longer wavelengths are used, and a $k$-correction applied, which calculates the magnitude a supernova would have had, were it observed in rest-frame $B$-band.  This correction is complicated by any colour variation that would change the relative magnitude at different wavelengths of the supernova's spectrum.  Colour variations occur both intrinsically, and due to dust absorption between us and the supernova.  Thus, the measurement of the standard candle is generally a simultaneous fit to these parameters, including colour, redshift, and light-curve width; sometimes the fit to cosmology is also done simultaneously.  

}

The original discovery of an accelerating universe was made using just 52 supernovae (10 with colour information, 42 in a single passband) at $z\leq0.83$.   The latest results all confirm the original discovery, now with in total over 1000 supernovae out to $z\sim1.9$, the most distant of which \citep{jones13} emitted the light we are now seeing 10.5\,Gyr ago.\footnote{Cosmologists usually define ages in the universe by referring to redshifts, not times, because redshifts are the observable quantity, while times are model dependent.  The light from an object at $z\sim1.9$ has been travelling for 10.5\,Gyr in an $(\omol)=(0.3,0.7)$ universe, but only for 8.4\,Gyr in an $(\omol)=(0.3,0.0)$ universe, or 7.4\,Gyr in an $(\omol)=(1.0,0.0)$ universe (with $H_0=70$\kmsmpc in all cases).  If you recall, this is how the discovery of the accelerating universe ($\oll>0$) solved the age problem, and allowed the universe to be older than the oldest stars. }    Type Ia supernova cosmology is now a very mature field, and small systematic errors that were negligible in the initial studies are now important, given the precision of the current data.  Over the last decade several different methods of determining the peak magnitude of supernovae have been trialled, and some small systematic biases were revealed.  None of these were anywhere near large enough to bring into doubt the detection of acceleration, but are important when trying to determine cosmological parameters to the few-percent level.

\subsection{Supernovae --- Velocities and Lensing }\label{sect:snlens}
\vspace{-3mm}

Supernova measurements are now so numerous that it has become feasible to look for features in the scatter about the magnitude-redshift relation.  What was once just treated as noise has become a signal.  
Scatter about the Hubble diagram occurs for a number of reasons.  Observational error and intrinsic diversity of the supernovae are the two least interesting for cosmology.  The two most interesting are peculiar velocities and lensing.  

{\bf Peculiar velocities} refer to any motion that is not just the smooth flow of the expansion of the universe.  They arise because galaxies tend to fall towards each other and form clusters and filaments.  These motions shift the redshifts away from what they would be if the galaxies were flowing perfectly with the expansion, and also shift the luminosities slightly due to Doppler beaming.  We will discuss this in more detail when we discuss measurements of the growth of structure using galaxy redshift surveys (Sect.~\ref{sect:growth}).  For the moment take it for granted that these peculiar velocities leave a characteristic imprint on the pattern of redshifts and magnitudes you would expect to see in your magnitdue redshift diagram.  This expected pattern depends on how clustered matter is and how fast galaxies fall towards these overdensities, all of which tells us about the strength of gravity on intermediate scales (between solar system and observable universe) and can be used to distinguish between different models of dark energy.  

The contribution of peculiar velocities is dominant at low redshift, because it comes in the form
\beq (1+\zo) = (1+\zbar)(1+\zp)\eeq
where the observed redshift, $\zo$, gets contributions from the cosmological redshift, $\zbar$, and the peculiar redshift $\zp$ due to the component of peculiar velocity along our line of sight (where the latter is calculated using the usual special relativistic formula relating velocities to redshifts).  Most supernova studies to date exclude all supernovae at $\zo<0.02$, because the expected peculiar velocity contribution below that redshift exceeds 5\%.  

The statistical properties of galaxy motions (and thus the supernovae within them) can be predicted using a matter power spectrum \citep{hui06,davis11}.  Velocities can be decomposed into mulitpoles, each of which quantifies the motion on a different angular scale.  The lowest multipole is the monopole -- the isotropic expansion or contraction.  The next multipole is the dipole -- a bulk flow in a single direction.  The higher order multipoles then contain more complex patterns.  Dipole detections using supernovae have been made by \cite{dai11,colin11} while \cite{haugbolle07} were able to detect up to the quadrupole. 

Peculiar velocity measurements made with large surveys of galaxy distances or the CMB have shown (some controversial) evidence of velocities that are larger than predicted in $\Lambda$CDM \citep{kashlinsky08,watkins09,feldman10,abate12,lavaux13}, making them one of the few observables that's putting tension on the standard model, but other similar measurements see no discrepancy \citep{nusser11,osborne11,turnbull12,ma13,planckInt_XIII}.  This is a very interesting diagnostic, because the speed at which matter falls into overdensities can distinguish between some modified gravity models that predict identical expansion histories.  However, it is a difficult measurement to make, and could be susceptible to systematic errors,\footnote{Peculiar velocities are detected by comparing the measured distance to the predicted distance for an object at that redshift.  A deviation from zero indicates a peculiar velocity.  So that means that any unidentified measurement error will indicate a non-zero velocity, even if the true velocity is zero.} so more study is warranted.  With future wide-field studies both of galaxies and supernovae being planned, peculiar velocities promise to be a very powerful probe in the near future \citep{koda14,johnson14}.

The second interesting feature that would cause scatter about the magnitude-redshift relation is {\bf gravitational lensing}, both strong and weak.  We discuss lensing more in Sect.~\ref{sect:stronglensing} and~\ref{sect:weaklensing}.  For supernovae the important feature for cosmology is mostly weak lensing.  

The light from a supernova reaches us after passing through the complex roller-coaster of gravitational potentials that are caused by the clumping of matter along the line of sight.  When you take a representative volume of the universe at the present day, you would find that most of the matter has fallen into overdensities (galaxies and clusters of galaxies) but that those overdensities only take up a small part of the total volume.  Most of the volume is taken up by underdensities, often called voids (where `void' doesn't imply complete emptiness, just an underdensity).  Thus most of the sightlines to supernovae will go through, on average, underdense regions.  This has the effect of slightly de-magnifying most supernovae.  Meanwhile, a few sightlines pass close to overdensities, and these can be strongly magnified.  This results in an asymmetric distribution of a few very bright supernovae and many more slightly faint supernovae.  The effect becomes more pronounced at higher redshifts, as the light has had longer to be lensed --- but don't worry, the total photon flux is conserved, so when you average these supernovae you still get a mean that is reliable (i.e. at the true non-lensed magnitude that you would expect for that redshift), so your magnitude-redshift diagram should still be accurate.  

This does highlight the fact that one should be wary of excluding obvious bright outliers from the magnitude-redshift diagram, because you need those occasional bright outliers to balance the many slightly faint supernovae.  Supernovae should only be excluded based on some property other than their magnitude.  For example, there are super-luminous supernovae that can legitimately be ruled out of the magnitude-redshift diagram because they are distinguishable by their peculiar spectra.  

To use lensed supernovae for cosmology the easiest thing to do is to look for the asymmetric distribution of scatter about the magnitude-redshift diagram.  The more complicated but much stronger diagnostic is to look for correlations between the supernova magnitude and the foreground density of galaxies.  Supernovae behind overdense regions should be brighter, on average, than those found behind empty regions.  This has recently been done and was found to be consistent with the $\Lambda$CDM model \citep{smith14}; and the technique has strong prospects for the future.

\newpage
\section{Cosmic Microwave Background}
\vspace{-3mm}

Undoubtedly the queen of cosmological observables is the cosmic microwave background (CMB).  This radiation was emitted when the universe was young and still relatively simple.  The physics of the early universe is just the interaction of temperature, density, and pressure of whatever components were around at the time (such as radiation, neutrinos, normal matter, dark matter, and dark energy).  The well known equations of thermodynamics and particle physics allow us to very precisely calculate how the components interact, and what we should observe in this radiation from the big bang.    

The first observation of the CMB by Penzias and Wilson in 1964 earned them the 1978 Nobel Prize for Physics because it was confirmation that the universe was once in a hot, dense state, and thus strongly supported the big bang theory.  It was a profound observation because it told us about the origin of our universe.  Until then the steady state theory\footnote{The steady state theory is one in which matter gets created as the universe expands, so that the density of the universe is constant and the universe is eternal.  This was philosophically attractive because it was based on the perfect cosmological principle, which states that we should not find ourselves at any special position in the universe, nor at any special time.  Since the universe evolves we have to fall back on the plain old ``cosmological principle'' which is simply that we are not in a special position, because we do find ourselves at a particular time.  One could argue that this is not a ``special'' time, and therefore the perfect cosmological principle still holds, but that is a debate for a different time.} had competed on almost equal ground with the big bang theory. 


When the universe was still hot and dense enough to be plasma, there were no atoms to put absorption or emission lines in spectra.  So the CMB remains to this day the most perfect black-body spectrum ever measured.  The temperature of that blackbody is $2.7255\pm0.0006$ K \citep{fixsen09}.  Interestingly, the most precise individual measurement of that temperature is still that made by the first of the CMB space telescopes, the COsmic Background Explorer (COBE) 
\citep{fixsen96}, since the space telescopes that followed only made differential temperature measurements and were not sensitive to the blackbody, by design.  The most precise overall measurement quoted above comes from recalibrating the COBE data by utilising velocity information from WMAP, and combines it with other balloon and ground based data \citep{fixsen09}.


COBE was also the first instrument to detect the fluctuations in the CMB, that represent the slight over- and under-densities in the early plasma of the universe that formed the seeds of structure we see today.  These temperature fluctuations are tiny --- on the order of $T\sim10^{-5}$K, which shows just how uniform the early universe really was.  Nevertheless, these fluctuations contained the overdense seeds that would eventually collapse under gravity to form the galaxies, filaments, and clusters that make up the cosmic web today.  

The early universe was a combination of various components, and the interaction between these components formed the pattern of density fluctuations that we see in the CMB.  It is one of the most astounding demonstrations of how well we understand the universe that we can take the equations for density, pressure, \& gravity that work for sound waves here on Earth, apply them to the early universe, and make predictions for what the radiation from that time should look like.\footnote{The fluctuations in the CMB were predicted by a series of papers in the 1970s \citep{harrison70, peebles70, zeldovich72, doroshkevich78}.} It's even more astonishing that humans were then able to design and build telescopes  to measure these fluctuations, launch those telescopes into space or float them around places like Antarctica by balloons and, lo-and-behold, detect the fluctuatons just as predicted.  

Predicting the patterns in the CMB requires some assumptions about the initial conditions.  The predictions usually start from random (gaussian) fluctuations, of the kind that you expect to exist if the universe began with an inflationary phase that converted quantum fluctuations into real fluctuations, which then seed structure.\footnote{The assumption of gaussian initial conditions has now weathered significant scrutiny.  Statistics that detect non-gaussianity, such as the three-point correlation function and bi-spectrum, are used to search for deviations from gaussianity.  Non-gaussianity does develop as structures evolve, but what would be really interesting is a discovery of primordial non-gaussianity, which would indicate that the simplest models of inflation would be in trouble.  So far, however, no deviations from gaussianity have been seen, with now quite stringent limits \citep{planck13_XXIV}.}  It's remarkable that, starting from completely random fluctuations, patterns develop.  These patterns arise due to the variety of ways in which different components of the universe interact, and due to the timescales imposed by the rate of expansion of the universe --- namely how far light could travel, and how quickly those components of the universe diluted and stopped interacting with each other.  

The components of the early universe interacted in various ways.  Dark matter essentially ignored everything else from the get-go, and simply collapsed under its own gravity without any opposing pressure forces.  Normal matter on the other hand had to deal with very high pressures, and thus initial density fluctuations propagated as sound waves.  For the majority of the time we're interested in, most of the normal matter existed in the form of protons and electrons.  The very high pressure was primarily generated by interactions with photons, whose mean-free-path was so short as to make the protons, electrons, and photons, act as a single fluid.  The CMB was emitted when the universe had expanded and cooled enough for electrons to stick to protons and form hydrogen.  With the electrons out of the way the mean-free-path of photons became essentially infinite.  Shortly after this time the sound waves froze, because the pressure supporting the wave dropped dramatically when the photon pressure disappeared. 

Meanwhile, neutrinos, which interact very weakly, had decoupled from the rest of the plasma at a very early time, and were free-streaming at close to the speed of light.  They wipe out small-scale fluctuations, because they tend to spread out the distribution of energy density from the initial fluctuations in which they began.  Dark energy had negligible effect at this early time, because its density was so low compared to the density of matter.  As a result, the CMB power spectrum alone does not give strong constraints on dark energy. 

{\blue 
We observe the CMB as an almost uniform bath of radiation arriving from all directions.  The fluctuations in the CMB are quantified by making an angular power spectrum, which quantifies correlations as a function of angular separation.  Basically, to make a power spectrum, choose a random point on the sky and see whether it is hotter or colder than average.  Then look at a point a certain separation, e.g.\ 1 degree, away from the original point and see whether that is also hotter or colder than average.  If both points are hot, or both points are cold, then there is a positive correlation.  Repeat for every point on the sky.  If that positive correlation tends to hold on average then you get a strong positive signal in your power spectrum at that scale (``high power'').  That's the power at a single scale, i.e.\ one point on the power spectrum.  To build the full spectrum you repeat the process for all possible separations on the sky.  If the pattern is completely random you will not see any structure in a plot of power vs scale.  However, if there are patterns, you will see peaks and troughs corresponding to preferred separations. 
}

\begin{figure}\sidecaption
\includegraphics[width=110mm]{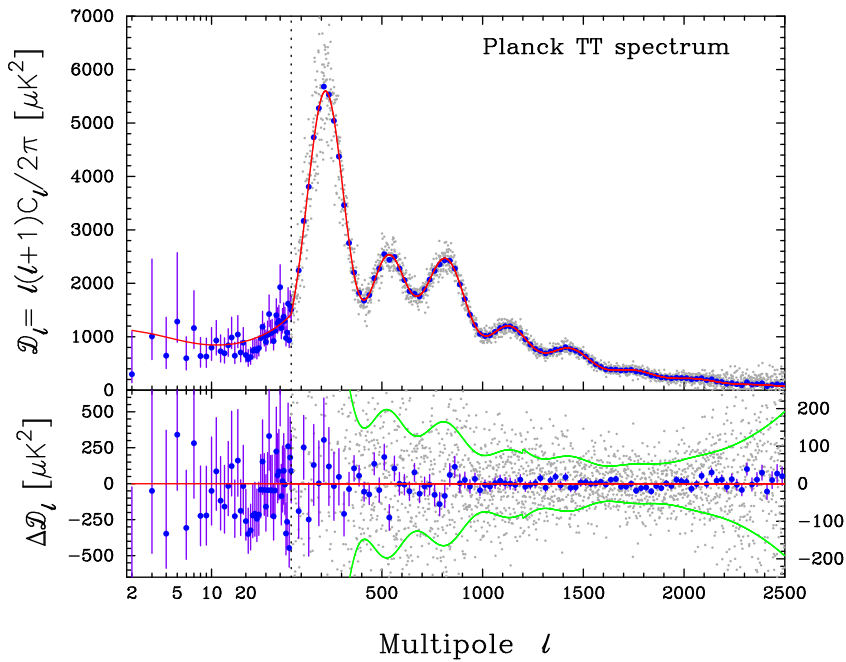} 
\caption{Most recent power spectrum measured by the Planck satellite (reproduction of Fig.~1 in \cite{planck13_XVI}), demonstrating an amazing match between theory and data.  This complex pattern is reproduced by only six primary parameters (although numerous nuisance parameters are used to remove foregrounds and other known non-cosmological effects, such as emission from our own galaxy). \vspace{2cm}}
\label{fig:PlanckPk}
\end{figure}

{\blue 
What we find when we perform this task, is that there appears to be a preferred separation of about 1 degree, where hot spots (and cold spots) are correlated (see Fig.~\ref{fig:PlanckPk}). Physically, this corresponds to the distance that sound waves could have travelled between the birth of the universe and the time when the CMB was emitted (the surface of last scattering).  The additional structure is closely related to the overtones, or harmonics, of the primary wavelength.  The second peak corresponds to the pattern of rarefractions, on the scale where a sound wave has had time to compress and rarefy once before last scattering.  The third peak corresponds to the second compression, and so on.  The spacing and height of these peaks were determined by several effects, one of the dominant ones being the ratio of baryonic matter to dark matter, another being the damping effect of radiation.  

It is remarkable that the complex pattern in the CMB (Fig.~\ref{fig:PlanckPk}) can be reproduced with only six primary parameters.  The first three relate to the contents of the universe --- the energy densities of {\em baryonic matter} ($\ob$), {\em cold dark matter} ($\oc$), usually expressed in `physical units' ($\obhh$ and $\ochh$, where $h=H_0/100\kmsmpc$); and {\em dark energy} in the form of a cosmological constant ($\oll$).\footnote{\blue The cosmological constant can be replaced by a parameter describing the apparent size of the sound horizon scale at last scattering, because dark energy does not strongly effect the pattern of fluctuations generated, but rather changes their apparent scale on the sky (because dark energy is negligible at early times, but has a large effect on the geometry and expansion history of the universe since then).}  
The other three relate to the initial conditions or conditions when the CMB is emitted.  Firstly, the {\em spectral index}, $n_s$, tells us the scale-dependence of initial fluctuations.\footnote{\blue If there are equal fluctuations on all scales, $n_s=1$; if there are more large-sized fluctuations than small-sized fluctuations, $n_s<1$.  If inflation was the process that generated the initial fluctuations then we expect to see $n_s\lsim 1$ (for most models) and the latest measurement finds $n_s = 0.9603 \pm 0.0073$ \citep{planck13_XVI}.}  Secondly, the amplitude of those initial fluctuations (given several different symbols, but often $A_s$), tells us how intense those fluctuations were.  And finally, the {\em optical depth} to reionisation, $\tau$, gives information about how much the photons are scattered during their travels, which depends on the ionisation state of the matter in the universe over time.}
  
A series of space, ballon-borne, and ground-based telescopes have measured the cosmic microwave background in intricate detail.  The space-based telescopes typically measure the large-scale correlations best, because they survey the entire sky, while ballon-borne and ground-based telescopes survey smaller patches of the sky in greater detail, and therefore have greater precision on small scales.  The three major space-telescopes have been:  COBE in the early 1990s \citep{smoot92,bennett96}; the Wilkinson Microwave Anisotropy Probe (WMAP) with results over the decade starting from 2003 \citep{bennett03,komatsu11,hinshaw13}; and Planck whose first results were released in 2013 \citep{planck13_XVI}. The first major balloon-based surveys were BOOMERanG \citep{debernardis00,melchiorri00} and Maxima \citep{torbet99,hanany00}.  More recently, the Atacama Cosmology Telescope \citep{dunkley11,sievers13}, and the South Pole Telescope \citep{keisler11}, have both produced superb data sets of small scale CMB fluctuations.

One of the most important measurements from the CMB gives an indication of the {\bf curvature of space}.  We know the wavelength of the first acoustic peak theoretically from our calculations of sound waves in the early universe.  So it can be used as a standard ruler.  By measuring how large it appears we can measure how far it is away.  The apparent size, however, depends on the curvature of the universe.  In a negatively curved universe the angles of triangles add up to less than 180$^\circ$, so the angular size of a standard ruler would appear smaller than in a positively curved universe, for which the angles of triangles add up to more than 180$^\circ$.  This is somewhat degenerate with the length of the standard ruler, which depends on quantities such as the matter density, and therefore this statement is model-dependent. 
However, the degeneracy can be broken by the smaller-scale peaks, so the position of the first peak in the CMB gives a strong indication of curvature.\footnote{Although in some models, such as $w$CDM the degeneracy between curvature and other parameters is stronger and thus the constraints are weaker.}  That in turn tells us whether the universe is close to the critical density --- the density that would make the universe flat. 

COBE was the first to measure the anisotropies, but did not have the resolution to measure the first acoustic peak.  So the first precise measurement of the flatness of the universe was determined by BOOMERanG and MAXIMA, who both found the universe to be close to flat \citep{debernardis00,melchiorri00,hanany00}.  

The second space telescope, the Wilkinson Microwave Anisotropy Probe (WMAP) dominated the CMB field during the decade beginning 2003.  The precise measurements of the smaller peaks confirmed to much greater precision the baryonic and dark matter densities, and the flatness of space.  In 2013 the first results from the latest space telescope, Planck, were released --- see Fig.~\ref{fig:PlanckPk}.  This is simply one of the most beautiful data sets I have ever seen.  The $C_\ell$ correspond to the power at multipole moment (inverse angular scale) $\ell = \pi/\theta$, where $\theta$ is the angular separation.  

There is an enormous amount of information in the CMB.  More than I will discuss here.  For example, you can correlate more than just the temperatures at different positions.  You can also consider the polarisation correlations, the effect of weak lensing, and the cross-correlations between temperature, polarisation, lensing, and foregrounds.  One of my favourite measurements is the polarisation seen in WMAP when you stack the data centred on hot or cold spots; see Fig.~5 and 6 of \cite{komatsu11}.  Since the polarisation depends on the velocity of gas, this measurement was able to detect the motion of the gas generating the sound waves, confirming the source of the temperature fluctuations.  



\vspace{-10mm}
\subsection{CMB foregrounds}
\vspace{-5mm}

Since the CMB is such a uniform background radiation, it also serves as a beautiful backlight for revealing matter distribution in the foreground.  The gas and galaxies it passes through en route to our telescopes interact with the CMB and change it in a variety of ways.  

{\bf The Sunyaev-Zeldovich (SZ) effect} occurs when  low-temperature CMB photons pass through the hot x-ray gas of a galaxy cluster, and are inverse Compton-scattered to higher temperatures.  This has been detected as the CMB radiation behind clusters appears hotter than average \citep{sunyaev72,ma13}. 
{\blue Interestingly, this effect is distance independent, because we're not looking at things that shine, but rather `holes' in the CMB sky --- the shadows of structure.  So the SZ effect can be used to discover very high redshift clusters of galaxies \citep{weller02}.  This can be used to do cosmology by counting clusters as a function of redshift.  If you recall, the number counts of galaxies were one of the early pre-supernova indications that the universe was accelerating.  Number vs redshift reveals cosmology because it relates to volume vs redshift, which depends on the cosmological model \citep{benson13,hasselfield13,planck13_XX}. }

Most importantly for dark energy, the {\bf Integrated Sachs-Wolfe (ISW) effect} is the gravitational redshifting that occurs as light travels through over- and under-densities.  In a flat, matter-only universe, the blueshift of light falling into a cluster potential is exactly cancelled by the redshift as it climbs out.  However, if dark energy exists, then that balance is broken.  The potential decays during the transit, so the light doesn't lose all the energy it gained on the way in, and there is a net blue-shifting of the photons.  Light behind very large clusters thus appears hotter than average. 
Since this only occurs if something --- such as dark energy --- is causing the potential to decay, this is a difficult observation \citep{fosalba03, boughn04,giannantonio08,ho08,planck13_XIX} to explain in theories that try to explain the acceleration without dark energy.\footnote{\blue Potentials also decay in a radiation dominated universe, or a matter dominated universe that is underdense; but the signature we see in the CMB is consistent with the prediction of dark energy.}  
   
\vspace{-2mm}
{\blue The plain old `Sachs Wolfe' effect, without the `integrated', refers to the primordial gravitational redshift photons emitted at the time of last scattering experienced if they were in density fluctuations.  
The hot and cold spots in the CMB don't uniformly map to overdensities and underdensities.  There are two effects at play.  Firstly, gas gets hot when it gets compressed.  So you would expect overdensities to be hot and therefore blueshifted compared to average.  However, when light is leaving an overdensity it has to climb out of a gravitational well.  Therefore it gets redshifted, and appears cooler than light from underdensities.  How do these two effects compete?  Well, the gravitational redshift effect is only large if you have very large overdensities.  So on large scales (small $\ell$ in the power spectrum) the gravitational redshifting is most important, and dense spots look cool.  Whereas small overdensities only have small gravitational redshifts, so their compression heating dominates the wavelengths we see.  So on small scales (large $\ell$ in the power spectrum) dense spots look hot.  The crossover occurs at approximately $10^{\degree}$.  

\vspace{-2mm}
The power spectrum on the very large scales is dominated by the SW and ISW effects, and the large scales are one place where slight anomalies from the standard cosmological model seem to appear, such as low power on the largest scales and some surprising alignments between multipoles.  Whether these are a statistical fluke, observational error (although low power and alignments are seen in both WMAP and Planck results), or something fundamental, has yet to be determined \citep{copi10,copi13}.
}

\begin{figure}\sidecaption
\resizebox{0.74\hsize}{!}{\includegraphics[width=84mm]{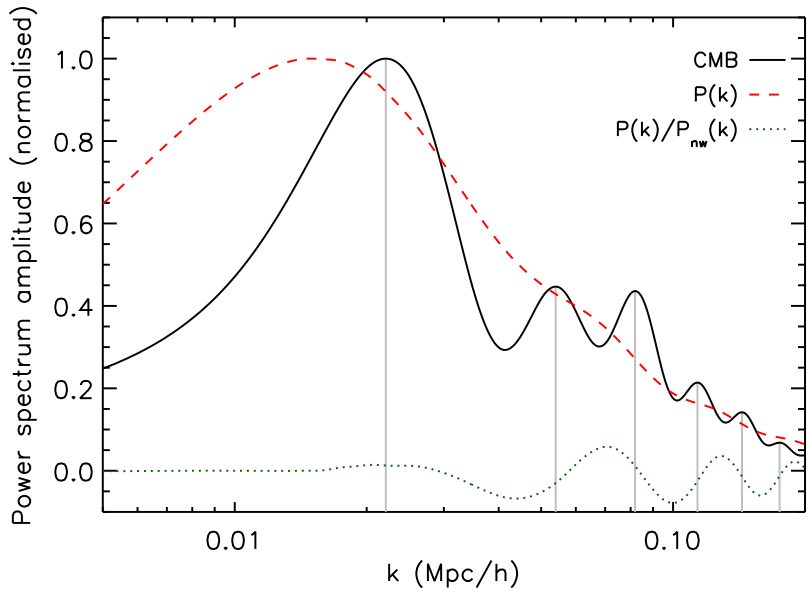}}
\caption{Comparison of the power spectrum of the cosmic microwave background at last scattering (black solid line with grey vertical lines marking the peaks), to the power spectrum of matter at the present day (red dashed line) for a flat $\Lambda$CDM model.  The green dotted line shows the matter power spectrum after dividing by a smooth no-wiggles curve to show the BAO positions more clearly.   The positions of the BAO peaks are out of phase with the CMB both because of velocity overshoot and projection effects.   
\vspace{1cm}}
\label{fig:cmbvsbao}
\end{figure}

\vspace{-5mm}
\section{Large Scale Structure (LSS)}

The seeds of structure that we see as fluctuations in the cosmic microwave background, gave birth to the large scale structure we see in the universe today.  That large-scale structure consists of galaxies, which appear in clusters and filaments, that make up what is known as the cosmic web.  Since the patterns in the CMB are not random, the positions of galaxies will not be random either.  The preferred separation of density fluctuations seen in the CMB should also appear as a preferred separation of the galaxies that eventually formed from those fluctuations.  

The fluctuations we see in the CMB will take billions of years to form galaxies at those positions, and it is important to note that the nearby galaxies we see don't correspond to any specific CMB fluctuations.  Rather the statistical pattern of galaxies should correspond to the statistical pattern seen in the CMB (if the universe really is homogeneous and isotropic on large scales, which observations show it to be \cite{scrimgeour12}).  A galaxy power spectrum can be created, in a similar way to the power spectrum created for the CMB, with the difference being that there is 3D information in the galaxy positions, whereas the CMB is just a spherical shell.  This 3D information becomes very important for measuring growth (Sect.~\ref{sect:growth}) and allows us to use radial separations to determine $H(z)$ (Sect.~\ref{sect:ap}).

The peaks in the CMB and BAO are not in exactly the same positions (see \cite{meiksin99} and Fig.~\ref{fig:cmbvsbao}).  For one thing, they are $\sim\pi/2$ out of phase due to velocity overshoot.  This occurs because when pressure stops driving the sound waves and they're `frozen in', the velocities of particles remain.  The scales that were in the process of becoming an overdensity are the ones into which the moving particles are falling fastest at last scattering, and therefore are the ones that initially grow most rapidly (see Fig.~8 of \cite{lineweaver97}).    Since velocities in a sound wave are maximal between density peaks and troughs, this results in the $\sim 90^{\degree}$ phase shift between CMB and BAO (see also Fig.~1 of \cite{meiksin99} and Fig.~6.9 of \cite{percival07proc}).  Another shift occurs because the sound waves stop slightly {\em after} the CMB is emitted, an effect that is important for the positions of Baryon Acoustic Oscillations (Sect.~\ref{sect:bao}).  In addition, projection effects arise because the CMB is measured on a spherical shell, while BAO are measured in 3D.

\begin{figure}\sidecaption
\resizebox{0.65\hsize}{!}{\includegraphics[width=84mm]{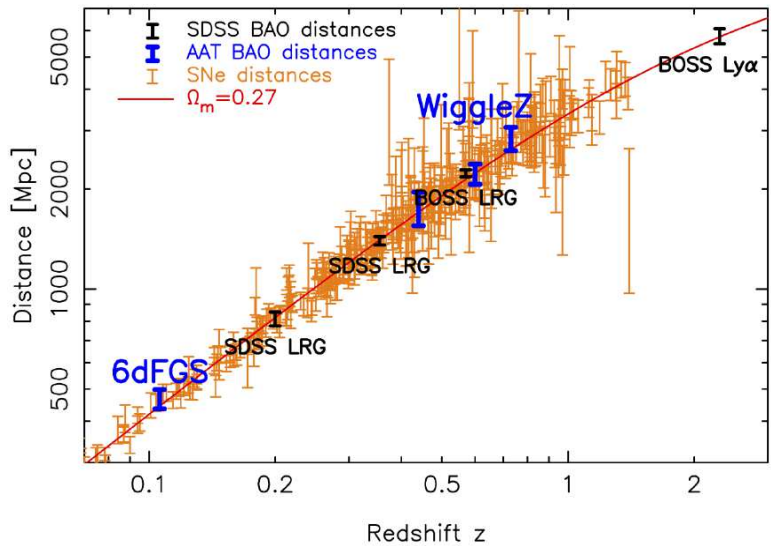}}
\caption{Magnitude-redshift diagram comparing supernova distances to BAO distances.  The supernovae are more numerous, and the BAO more precise.   Combined they give strong and complementary constraints on the expansion history of the universe.  The BOSS Lyman-$\alpha$ detection is the most distant distance measurement to date (excluding the CMB).  Figure, adapted from Fig.~12 of \cite{blake11bao2} by Chris Blake. \vspace{2cm}}
\label{fig:bao_vs_sne}
\end{figure}

\newpage
\subsection{LSS --- Baryon Acoustic Oscillations}\label{sect:bao}
\vspace{-5mm}

Baryon acoustic oscillations (BAO) refer to the preferred galaxy separation that arose due to sound waves (acoustic oscillations) in the baryonic matter (not dark matter) of the early universe.  Recall these oscillations occurred in the early universe, but the term BAO usually refers to the detection of that preferred separation in the distribution of galaxies.  Again, it can be used as a standard ruler, to measure distance.

The size of the BAO standard ruler is slightly longer than it appears in the CMB.  Recall that the CMB was emitted when the universe expanded enough that most photons had space to propagate freely.  Recall also that the photons were providing the pressure to drive the sound waves in the baryons, so without photon pressure, the sound waves should stop propagating.   However, after the CMB was emitted (the surface of last scattering) the sound waves did continue for a short while (known as the drag epoch), because the photons were so much more numerous than the baryons.  (about a billion to one).  Even though most of the photons were able to propagate freely there were still enough to continue interacting with the baryons and drive the pressure of the sound waves for a little longer.  Thus the CMB was emitted at $z\sim1100$ but the BAO scale didn't freeze in until about $z\sim1060$,\footnote{Note that a common approximation for the redshift that marks the end of the drag epoch is \cite{eisenstein98} Eq.~4, which gives approximately $z_d=1020$ for the best fitting $\Lambda$CDM model.} making the standard ruler of the BAO about 4\% larger than the standard ruler in the CMB. 

After the sound waves were frozen, the compressions started collapsing under their own gravity and formed overdensities at the characteristic separation of that initial sound wave.  For an excellent description of the formation of the BAO see Dan Eisenstein's website\footnote{\url{https://www.cfa.harvard.edu/~deisenst/acousticpeak/acoustic_physics.html}}.  

One thing that makes the BAO such a reliable cosmological probe is that they are on such large scales ($\sim150$Mpc) that astrophysical processes can not really effect them.  They shift around a little due to peculiar velocities, but even that can be taken into account, because if you know where the overdensities are you know how fast galaxies should be moving towards them, and can remove the effect.  This is known as reconstruction \citep{eisenstein07reconstruction,padmanabhan12,kazin14}. 

The presence of the BAO is very subtle.  The overdensity we measure is less than a 1\% deviation from the homogeneity of the universe \citep{scrimgeour12}.  It can't be seen in any individual galaxy separations.  It needs hundreds of thousands of galaxies, mapped in a large 3D volume of space, to be noticeable.

The BAO have now been measured by several groups making large 3D maps of the distribution of galaxies in the universe.  These include the 2 degree field galaxy redshift survey (2dFGRS), the Sloan Digital Sky Survey (SDSS), the WiggleZ dark energy survey, and the Baryon Oscillation Spectroscopic Survey (BOSS) \citep{cole05,eisenstein05,percival07,percival10, blake11bao1,blake11bao2,anderson12,padmanabhan12,sanchez12}.  
Early BAO measurements averaged the measured galaxy separations over all directions in order to get sufficient signal-to-noise to detect the BAO peak.  Most BAO measurements now are done while preserving the angle information, i.e.\ the angle of the separations with respect to the line of sight \citep{gaztanaga09,kazin10b,sanchez13,kazin13}.  This gives important extra information, which allows you to measure features such as $H(z)$ (see Sect.~\ref{sect:ap}). 

In a recent exciting result the BAO were also detected using Lyman-$\alpha$ absorption at $z\sim2.3$.  This works by using quasars (active galaxies) as backlights, and detecting the absorption due to fairly diffuse neutral hydrogen clouds along the line of sight.  These hydrogen clouds should also be distributed in the typical large-scale structure pattern, and the BAO signature in these clouds was detected by the BOSS team in 2013 \citep{busca13,slosar13}.  Because the source quasars are so bright and can be seen so far away, this allows the BAO to be measured to much greater distances than can easily be done with galaxies themselves (see Fig.~\ref{fig:bao_vs_sne}).  All these measurements confirmed the supernova discovery that the expansion of the universe is accelerating, and that the dark energy is consistent with a cosmological constant.

For all their excitement, BAO are just the tip of the iceberg for what large-scale structure surveys can tell us.  The BAO when used as a standard ruler essentially give us the same information as the supernovae --- they tell us the expansion history of the universe (distance vs redshift), albeit in terms of angular diameter distance rather than luminosity distance (see Fig.~\ref{fig:bao_vs_sne}).  However, with the three-dimensional information in the distribution of galaxies we can learn a lot more.

\begin{figure}
\includegraphics[width=73mm]{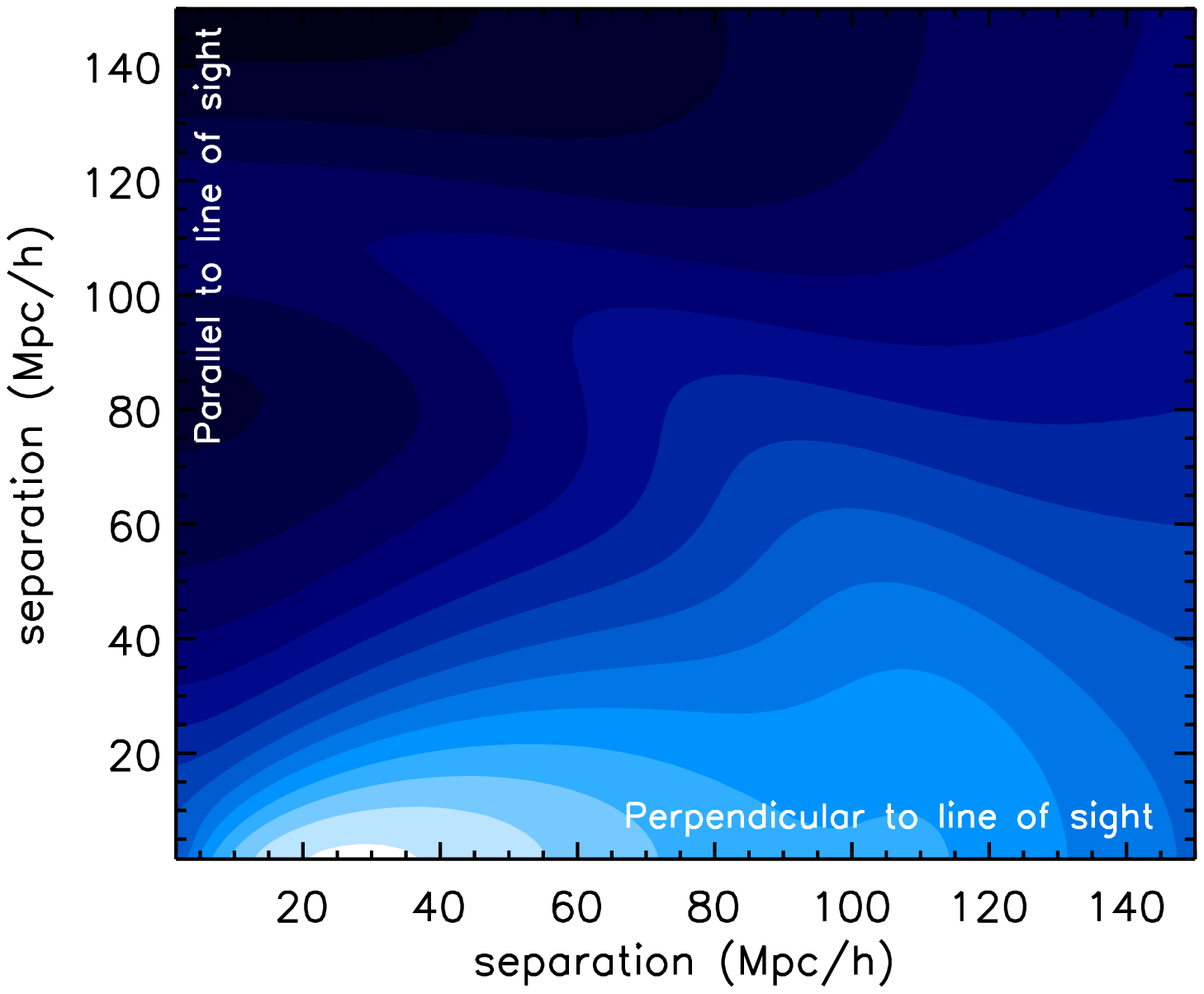}
\includegraphics[width=80mm]{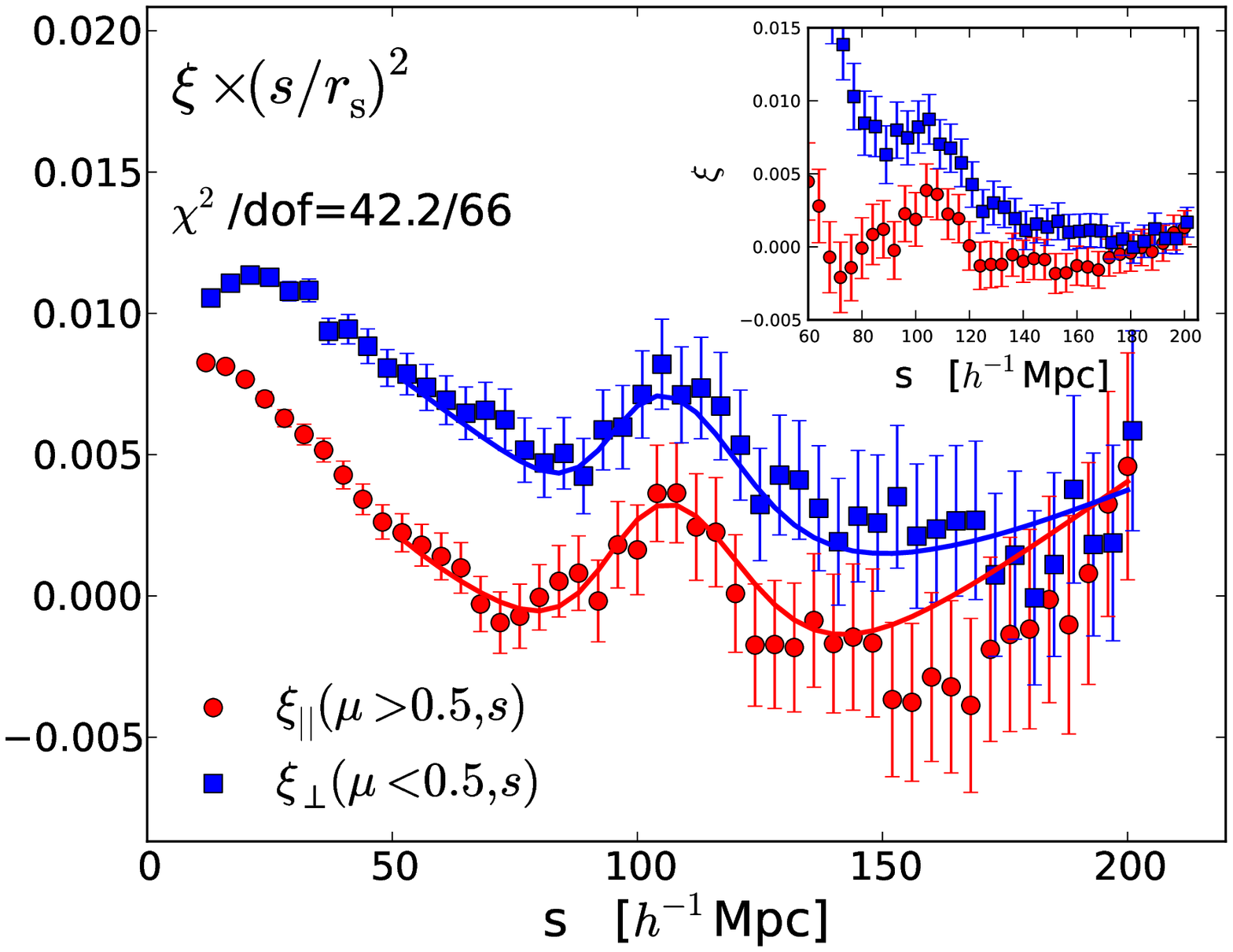}
\caption{Correlation functions of large scale structure.  Left panel: a model correlation function in 2D, considering separations as a function of angle.  The arc at approximately 110$h^{-1}$Mpc corresponds to the BAO peak, while the compression along the line of sight at smaller scales arises because of the growth of structure.   Right panel: Reproduces Fig.~1 of \cite{kazin13}, with BOSS data, where the red points correspond to the mean of the data in the wedge closest to parallel to the line of sight, while the blue points are the mean in the other half, closest to perpendicular to the line of sight. }
\label{fig:spheres}
\end{figure}

\vspace{-5mm}
\subsection{LSS --- Growth}\label{sect:growth}
\vspace{-5mm}

Sometimes knowing the expansion history is not enough to distinguish between possible models of dark energy.  Many models of dark energy predict the same expansion history (after all, that's the observation they were designed to explain!).  So to distinguish between them we need to use different {\em types} of observables.  Perhaps primary amongst these at the moment are measurements of the growth rate of structure. 

We know that gravity is attractive on the scale of our solar system, and our galaxy.  But we know that it appears repulsive on the scale of the universe as a whole.  So what happens in between?  Galaxy clusters are an intermediate scale, on which we would like to measure the strength of gravity, because here we can distinguish between different theories that predict the same expansion rate, but different growth histories.  

Growth can be measured 
using {\em redshift-space distortions}, by which we see the effect of galaxy velocities as they fall into clusters.  These distortions arise because we estimate the distance to the galaxy by its redshift, assuming the galaxies are in the Hubble flow.  However, in general, galaxies are not simply moving with the expansion of the universe, they are affected by each other's mutual gravitational attraction, so galaxies tend to be collapsing towards each other, forming clusters.  When this happens, the galaxies on the near side of the cluster are actually moving away from us slightly faster than they would if they were just in the Hubble expansion, so they get an extra redshift due to their peculiar velocity.  This extra redshift means that they appear to be slightly further away then they actually are.  On the other hand the galaxies on the far side of the cluster are falling towards us, and the extra blueshift they get from that infall makes them look slightly closer to us than they should.  This means that on large scales clusters appear compressed along the line of sight (Fig.~\ref{fig:spheres}).  When you measure many clusters you can statistically detect this flattening.  The more flat, the faster the infall.   This feature can potentially distinguish between different theories of gravity and different proposed types of dark energy.
   
Note that this is not to be confused with another redshift-space distortion effect, known as the {\em finger of god}.  The above infall only continues while the galaxies are falling towards the centre of the cluster.  Once a cluster has formed and virialised, you actually get the opposite effect, with the orbital motions seemingly lengthening the cluster along the line of sight.  This is known as the finger of god because in early galaxy surveys it appeared that every cluster had a preferred elongation that was pointing directly at us.  It didn't take long before the source of that elongation was elucidated \citep{kaiser87}.  

Many teams have now measured the growth of structure, including SDSS, WiggleZ, and BOSS \citep{gaztanaga09,blake11growth,reid12}.  This was a test that the standard $\Lambda$CDM model  could have failed --- several other theories predict different growth rates.  However, all measurements so far are consistent with $\Lambda$CDM, and strongly disfavour theories like DGP \citep{dgp00}, which preferred a slower rate of growth. 

{\blue
{\bf Observational considerations}

There is an important degeneracy in most of these measurements.  The growth rate is defined to be $f=d\log \delta/d\log a$, where $\delta = (\rho-\bar{\rho})/\bar{\rho}$, with $\rho$ as density and an overbar representing the mean.  This growth rate is tightly connected to the amount of clustering --- how strongly galaxies are attracted to a gravitational potential depends on the depth of that potential.  The strength of clustering at the present day is quantified by the parameter $\sigma_8$.  To derive $\sigma_8$ take an imaginary sphere of $8$ Mpc.  Place it at a random location in the universe and measure the density.  Repeat for many random locations and measure many densities.  The dispersion in densities about the mean is $\sigma_8$.  The choice of $8$ Mpc is somewhat arbitrary, but was chosen because when this was first being proposed, 8 Mpc was approximately the scale on which the dispersion in $\delta$ was of order unity.  These days with better measurements we know that the actual value is closer to $\sigma_8\sim0.83$ \citep{planck13_XVI}.  The expected value of $\sigma_8$ can be precisely predicted from the amplitude of fluctuations in the CMB {\em given a cosmological model}, but without using that theoretical input, what is usually measured is the product $f\sigma_8$.

A second complication arises because galaxies are not perfect tracers of the total matter distribution.  The growth of structure arises because of the total matter and energy density of the cosmos, but the galaxies are just the baryonic `icing on the cake', carrying only about $\sim5$\% of the total energy density themselves.  Like pebbles on a rough sandy beach, they'll gather in the deep potential wells, and less so on the peaks.  Thus galaxies give a biased view of the underlying matter distribution.  Different types of galaxies have different amounts of {\bf bias}.  Luminous red galaxies are the elliptical galaxies that are usually found at the centre of galaxy clusters.  Therefore they are strongly biased.  Blue galaxies are still star forming, and tend to be found on the outskirts of clusters and in the field.  Therefore they are more faithful tracers of the true matter distribution but less sensitive tracers of the highest-mass regions.  

The biggest modelling difficulty for large scale structure is the effect of {\bf  non-linear structure growth}.  While structures are small enough that perturbation theory does a good job at predicting their evolution, the models can be easily compared to the data.  However, when structures start to become strongly clustered, perturbation theory breaks down and analytical predictions fail.  Numerical simulations that calculate the full complicated evolution of growing structures are then needed.  
The limiting factor in large-scale structure measurements at the moment is not the observations, but rather the theory.  We are unable to reliably predict the small scale structures, and therefore throw out a lot of small scale data from our galaxy surveys.    High-redshift and less-biased galaxies are therefore useful, because they are closer to the linear theory prediction. 

}


\vspace{-5mm}
\subsection{LSS --- $H(z)$, Standard Spheres, 2D BAO, and Topology}\label{sect:ap}
\vspace{-5mm}

Growth is not the only reason that spherical features may appear oblate.  Another is known as the Alcock-Paczynski (AP) effect.  We measure distances to galaxies through their redshift.  In order to convert redshift to distance one needs to use a cosmological model.\footnote{Recall that this was how we figured out which model best matched the supernova data --- different models predicted different distance-redshift relationships.}  So when we map the distribution of galaxies, we typically use a fiducial model to convert redshift to distance.  This fiducial model leaves an imprint on what we call the `data'.  When we compare that data to other cosmological models what we actually do is compare the {\em ratio} of the test model distance to the fiducial model distance.  As long as the fiducial model is close to the true model this has been shown to be unbiased. 

However, the distance-redshift conversion for separations along the line of sight is different to the calculation for separations perpendicular to the line of sight.   So if you use the incorrect cosmological model to calculate distances, you will see things that are supposed to be spherical appear oblate or prolate.  The warping has a different functional form to the growth of structure, and therefore the two can be distinguished (although they are correlated) \citep{blake11ap}.  

Excitingly, the separations along the line of sight are related to $c\Delta{z}/H(z)$.\footnote{The distance along the line of sight appears in Hubble's law $v=HD$.  The BAO scale is small enough that the velocity difference between the galaxies can be approximated by $\Delta v=c\Delta z$, so $\Delta D=c\Delta z/H$, or $H(z)=c\Delta z / \Delta D$.}  So by measuring separations along the line of sight, we can measure the Hubble parameter at that redshift, $H(z)$, and thus measure the expansion rate directly (Fig.~\ref{fig:spheres}).
There are two features that have been used as standard spheres.  The first is the {\em shape} of the power spectrum or correlation function.  However, unless you can distinguish features in the shape, it only allows you to measure the ratio of the line of sight to perpendicular distances.  When you have measurements on sufficiently large scales to detect the BAO, you can use the BAO as a standard spherical shell.  This is very powerful, because the ruler has a known length --- the sphere has a known radius --- so if you can detect the BAO peak along the line of sight you can measure $H(z)$ alone.  
Again, this test measuring $H(z)$ is one that $\Lambda$CDM passed with flying colours. 

\begin{figure}\sidecaption
\resizebox{0.7\hsize}{!}{\includegraphics[width=120mm]{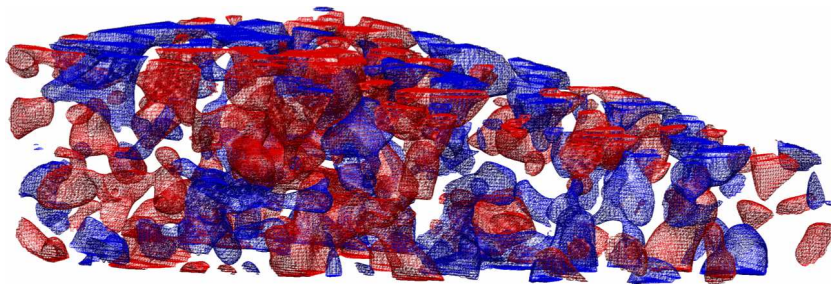}}
\caption{\blue Topology of one of the WiggleZ regions, ranging from $z=0.3$---$0.9$ smoothed on a scale of 20 $h^{-1}$Mpc.  Surfaces of constant density are shown, where red regions enclose the densest 20\%, and blue regions enclose the least dense 20\%, of the volume within the field. Reproduction of Fig.~1 of \cite{blake14topo} created by Berian James. }
\label{fig:topology}
\end{figure}


{\blue 
The latest method applied to large scale structure is a measure of topology.  Not to be confused with the topology of space, topology in this context refers to the connectedness of clusters and filaments of galaxies. The topology can be measured from a galaxy survey by smoothing the galaxy distribution, and determining surfaces of equal density (Fig.~\ref{fig:topology}).  This can be quantified using {\em Minkowski functionals}, which describe the (a) volume enclosed, (b) surface area, (c) curvature, and (d) ÔconnectivityÕ of the surfaces.
A common measure of connectedness is the genus statistic, which is roughly defined as the number of holes through something.  So a sphere has genus 0, a doughnut and a coffee cup both have genus 1, while most humans have at least genus 4 (but likely more depending on piercings).  

Topology is immune to some of the uncertainties that appear in other methods of quantifying structure, because any process that modifies the density field but not the rank-ordering of densities preserves the topology.  So topology is only weakly altered by non-linear structure formation.  Topological measurements provide a standard ruler once the shape of the underlying power spectrum is known.   Blake et al. \citep{blake14topo} used this to refine the WiggleZ results, and deliver measurements of distance vs redshift with a factor of two more precision (albeit with slightly stronger assumptions, i.e. knowledge of the shape of the underlying power spectrum).  }

\vspace{-5mm}
\subsection{\blue LSS --- Homogeneity and isotropy}
\vspace{-5mm}

{\blue One of the primary assumptions upon which most of the above measurements are based is that the universe is on average homogeneous and isotropic.  It is important to test whether this is true, and on what scales; not least because several theories invoke the presence of inhomogeneities as a way to explain the acceleration of the universe \citep{wiltshire07,ishak08,rasanen11}.  

There is strong evidence for the isotropy of the universe, in the form of the CMB, which appears the same from every direction \citep{fixsen96}.  However, to measure homogeneity (and even isotropy on smaller scales) one needs to look at a volume of space, for which one needs galaxy surveys. 

One way to test the homogeneity of a set of galaxies is to find the average number $N(< r)$ of neighbours for any given galaxy, up to a maximum distance $r$.  In an homogeneous distribution that should scale with the volume, so in three dimensions $N(<r)\propto r^3$.  However, if the sample is clustered then there will be an excess of  neighbours, and the distribution of $N(<r)$ will not scale with volume.  We know that the universe is clustered on small scales, so to look for homogeneity one can start with small spheres and search for the radius above which the number density scales with volume. This was tested by \cite{hogg05} for SDSS and by \cite{scrimgeour12} for the WiggleZ dark energy survey,
who find that the universe is within 1\% of homogeneous for $\gsim$80 $h^{-1}$Mpc, and that the data do not support significantly fractal models.  
}

\vspace{-5mm}
\section{Lensing}
\vspace{-5mm}

The final observational technique I will discuss in any detail is gravitational lensing. 
The bending of starlight around the sun during an eclipse was one of the first confirmations of general relativity.  These days the lenses we use for cosmology are not stars, but galaxies and clusters of galaxies.  Background galaxies are warped, magnified, and shifted in position, by matter in the foreground.  

\vspace{-5mm}
\subsection{Lensing --- Strong}\label{sect:stronglensing}
\vspace{-5mm}

Strong lensing occurs when a foreground galaxy or cluster (the lens) and a background galaxy (the source) are well enough aligned for multiple images to be seen --- the light from a single galaxy arrives from more than one direction.  Measurements of the multiple images can facilitate detailed mass models of the lens.  This is very useful for comparison with other means of measuring mass (e.g.\ measurements of virial motion or x-rays).  

One of the most powerful strong-lensing techniques is lensing time-delays.  If the source is variable, e.g.\ an active galaxy, then measuring the time delay between variations for two or more different images of it, gives the difference in path length that the light has travelled, which in turn allows one to
constrain cosmological parameters, such as the Hubble parameter $H_0$ \citep{refsdal64}.   Combined with CMB data from Planck, this test also gives answers consistent with standard $\Lambda$CDM (see Sect.~4 of \cite{suyu13}).  Once you allow the equation of state of dark energy to vary, the preferred value of the parameters tends towards a high Hubble constant and an equation of state of dark energy that is less than $-1$.  Some fuss has been made about this recently, but this is driven by the Planck data, which when unconstrained by other data sets tends to prefer very high Hubble constants and low $w$ (see lowest row of Fig.~12 in \cite{planck13_XVI}).

\begin{figure}
\includegraphics[width=80mm]{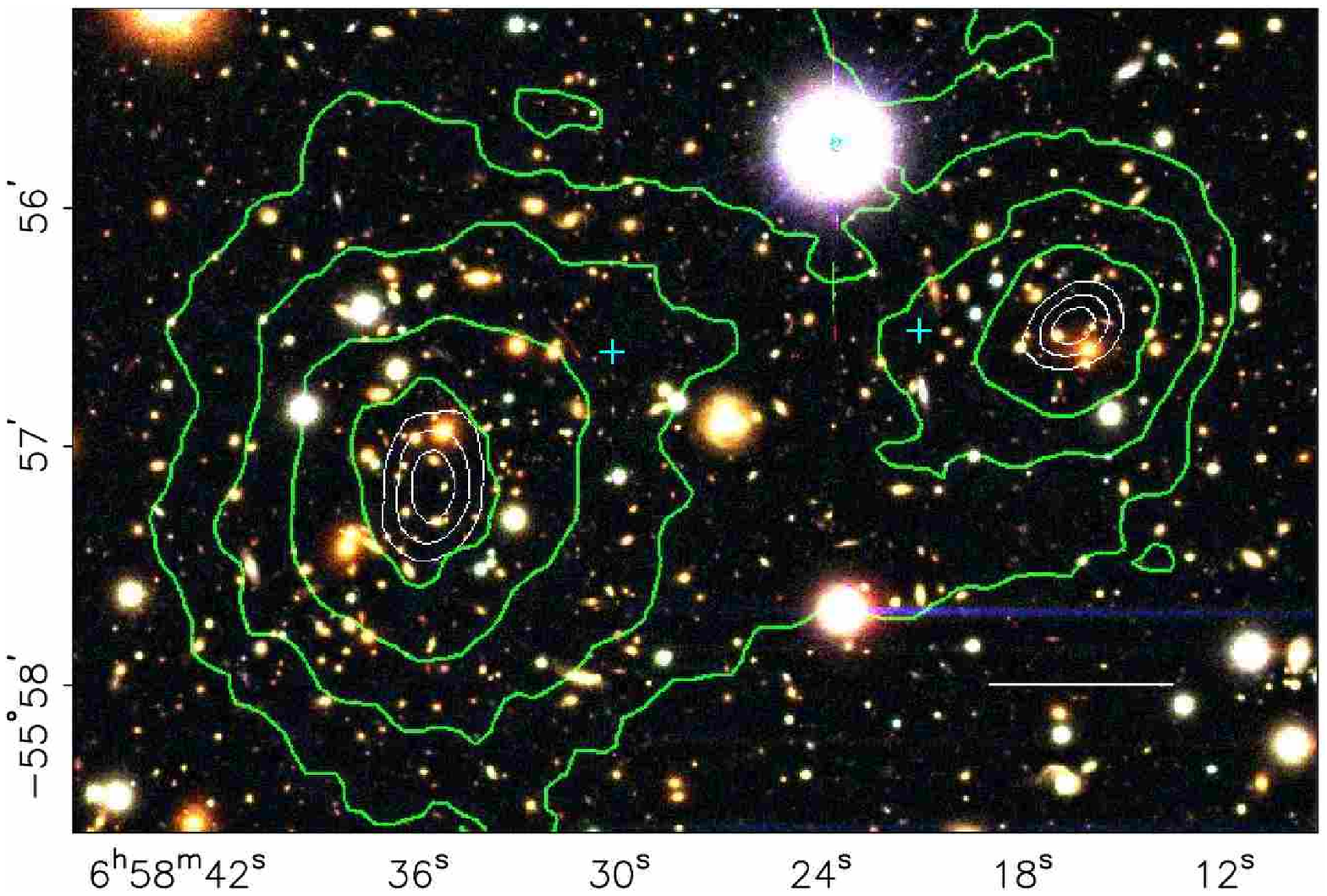}
\includegraphics[width=80mm]{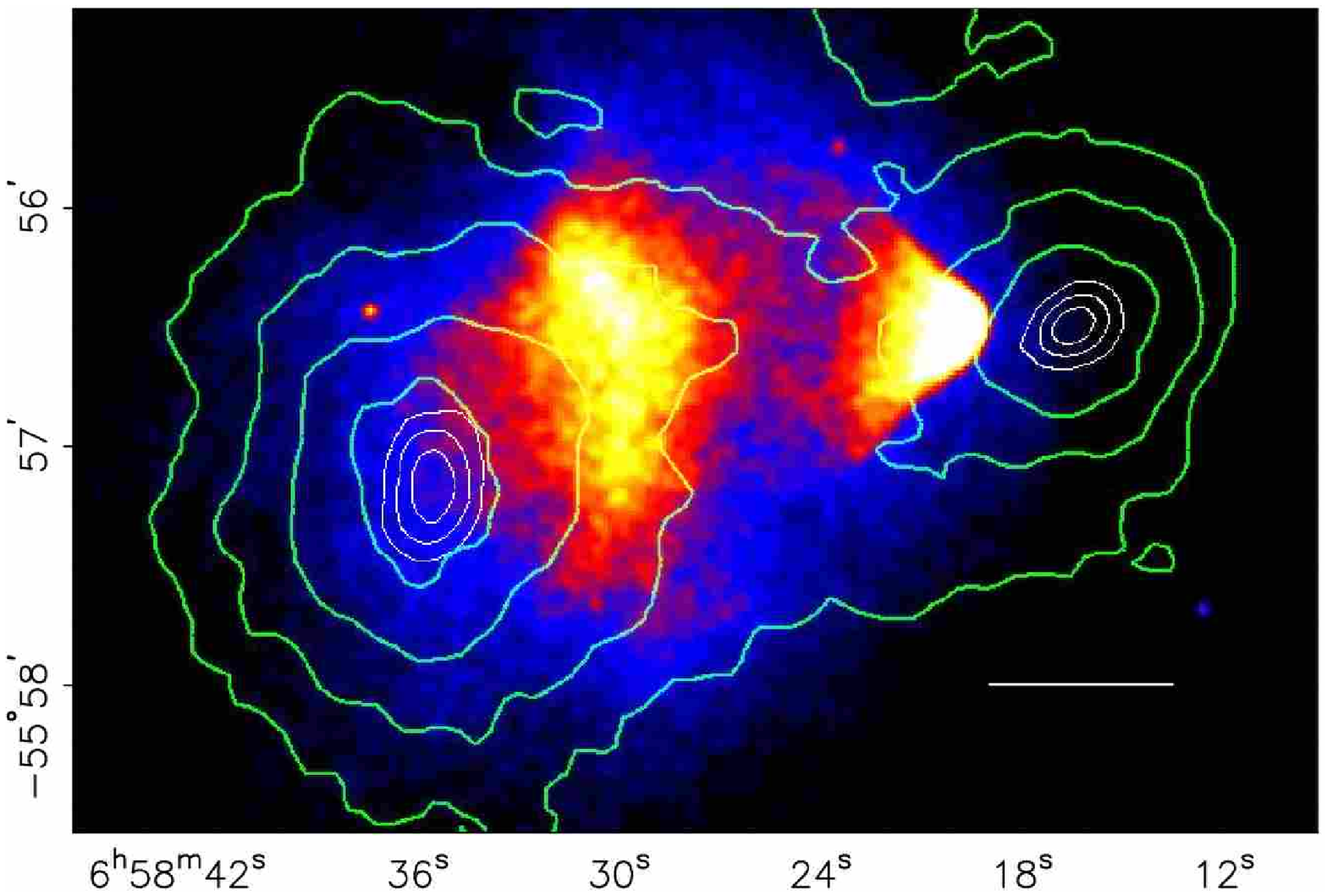}
\caption{Images of two clusters merging.  In both panels the green contours show the weak lensing determination of the mass distribution.  On the left the mass contours are superimposed on the distribution of galaxies --- the mass follows the galaxies fairly closely.  On the right panel the mass contours are superimposed on the x-ray emission from hot gas --- the gas has been slowed down as the clusters passed through each other, and thus is dragged away from the centre of mass.  This indicates that dark matter is collisionless, like galaxies.  (Reproduction of Fig.~1 in \cite{clowe06}.)}
\label{fig:bullet}
\end{figure}

\vspace{-5mm}
\subsection{Lensing --- Weak}\label{sect:weaklensing}
\vspace{-5mm}

Weak lensing occurs for all galaxies, as their light rides the rollercoaster of curved space en route to our telescopes.    We've already seen that the signal of weak lensing has been detected in supernova magnifications (Sect.~\ref{sect:snlens}).  In the cosmological context weak lensing usually refers to the warping of galaxy shapes, which has the potential to be an extremely strong cosmological probe.  

It is difficult to detect for an individual galaxy because the effect is usually small ($\sim 1\%$) and the intrinsic shape of any particular galaxy is not known well enough. 
Weak lensing can instead be be detected statistically in correlations of many galaxies (or in the CMB).  If we see that galaxies don't appear randomly oriented we can identify the presence (or absence) of foreground structure because that structure warps galaxy shapes in a correlated manner.  Lensing is therefore a very powerful probe because it traces the underlying density distribution directly, without needing biased tracers such as galaxy positions, to figure out where the overdensities are.
The correlations between shapes can be measured simply using an angular power spectrum, but the most powerful measurement of correlations is one that includes redshift information, called {\em tomography}.  

Detections of weak lensing include those made in HST COSMOS \citep{massey07}, SDSS \citep{lin12,huff12} and the CFHT Lensing Survey \citep{heymans12,benjamin13,kilbinger13}.  Measuring galaxy shears is technically very challenging, not least because the atmosphere and instrumental effects make intrinsic shapes difficult to measure, and alignments may occur for reasons other than lensing.  Major efforts are being made to understand and control systematics \citep{mandelbaum13}.

Some of the most powerful weak lensing results so far have been in measuring the mass distribution of specific clusters.  The most famous is the measurement of the mass distribution of the bullet cluster, which shows that the dark matter is coincident with the galaxies, but offset from the gas \citep{clowe06}.  This is a case where two clusters collided and passed through each other.  In that process the dark matter, which doesn't interact except by gravity, passed through smoothly, as did galaxies (the space between galaxies is so large that there are few direct collisions even when two large clusters collide).  However, the intergalactic gas was slowed by pressure and thus lags behind the dark matter (see Fig.~\ref{fig:bullet}).  This is one of the strongest indications that dark matter is really a collisionless particle, and makes it very difficult to explain by a modified theory of gravity. 
 
Cross-correlation between lensing results and estimates of the density distribution have started to appear, and already give constraints on modified theories of gravity \citep{bean10,simpson13}.  This combination of data can be particularly revealing, because the lensing potential depends on both the spatial and temporal parts of the metric, whereas clustering depends only on the spatial part.  Thus lensing-galaxy cross-correlations can constrain models that allow the temporal and spatial parts to differ (in general relativity they are the same).  For much more detail about weak lensing surveys see \cite{weinberg13}.

{\blue
{\bf Observational considerations}

Measuring galaxy shears is technically very challenging, because the atmosphere and instrumental effects make intrinsic shapes difficult to measure, and alignments may occur for reasons other than lensing.  One also needs knowledge of the redshift distribution and the matter power spectrum, both of which can have systematic errors.  Luckily, measurements of shape correlations have some built-in systematic error checks.  You only expect E-mode polarisation (circular or radial -- like wheels and spokes of a bike), not B-mode (spiral -- like the blades of a fan or turbine).  A positive detection of B-mode polarisation can be used to identify systematic error, including those you might expect from contamination due to intrinsic alignments of galaxies (galaxy orientations may not actually be intrinsically random, since they form along filaments and are subject to mergers and tidal torques). 
}

{\blue
\vspace{-5mm}
\section{Standard clocks}
\vspace{-5mm}

We've already talked about using standard candles, and standard rulers, to measure the expansion of the universe.  The final leg of the tripod would be to use standard clocks.  Anything of known duration, or age, can be used as a standard clock. 

\vspace{-5mm}
\subsection{Supernovae --- Standard clocks}
\vspace{-5mm}

Supernovae once again are handy, because as they are explosions of known duration, and with them we can measure something different --- time dilation.  

In an expanding universe, processes occurring in distant galaxies, which are moving away from us, should suffer time dilation. When time dilation is in action supernovae at redshift $z$ should take $(1+z)$ times longer to brighten and fade than their local counterparts.\footnote{Note this is not the special relativistic time-dilation formula, see appendix of \cite{blondin08} for a discussion.}  Observations both of light curves \citep{leibundgut96,goldhaber01} and time-series spectra \citep{riess97,foley05,blondin08}, have confirmed this $(1+z)$ time dilation, as expected. 

This has little bearing on dark energy, but confirms one of the absolute fundamentals of our entire cosmological model -- that the universe is expanding.  This may not seem like it needs confirming, but it seems common amongst the non-professional community to question the reality of the expansion of the universe.  Propositions such as ``tired light'' attribute redshift to some process other than expansion of space and recession velocities.  However, most such theories would not also predict time-dilation.  Therefore by using supernovae as standard clocks, and measuring time dilation, we have ruled out such theories.

\vspace{-5mm}
\subsection{Galaxy ages --- Cosmic chronometers}
\vspace{-5mm}

One other way to measure the expansion history of the universe is to use the average age of galaxies vs redshift.  The differential age evolution as a function of time, gives another measure of the expansion.  Galaxy ages are measured by modelling their spectra using stellar population models (their spectra are a sum of the spectra from all the stars).  Although absolute ages are difficult to measure, this technique gives you an {\em age difference} between galaxies at a range of redshifts, which again tells you how much expansion has occurred between those redshifts \citep{jimenez02,moresco12}.  One of the largest uncertainties measuring age differences comes from the difficulty in measuring galaxy metallicity, which is degenerate with age; and this is one reason these constraints are often not included in combined analyses for precision cosmological parameters.

\vspace{-5mm}
\subsection{Timing of structure formation}\label{sect:timing}
\vspace{-5mm}

There is yet another clock that we can use to inform us about what the universe is made of, and this is one of the simplest of all --- when the first structures formed.  We know how quickly matter falls under the influence of gravity, so we know how quickly density fluctuations should grow.  We know of several quasars out at $z>6$, which means they formed in the first billion years after the big bang.  Winding the clock backwards we can ask how big the fluctuations in the CMB had to have been, in order for such large structures to form so quickly.  The answer is that the CMB fluctuations needed to be on the order of a few percent --- far greater than the $10^{-5}$ fluctuations that we observe.  Thankfully there is a simple resolution to this conundrum --- dark matter.  While normal matter was pressure supported, dark matter had already started collapsing.  As soon as matter density dominated over radiation density, dark matter structures would have started to grow, and this gives enough time for the earliest structures we see to form.  This effect would be difficult to mimic with a modified gravity explanation of dark matter.
}

\newpage
\vspace{-5mm}
\section{The steak knives}
\vspace{-5mm}

And there's more!  I don't have space in this review to cover in any detail the many other techniques that inform our understanding of the properties of dark energy, but the list continues.
I haven't yet mentioned one of the earliest confirmations of the big bang theory --- {\bf nucleosynthesis}.  The theory of how the elements were created in the big bang, and in the proportions that we see them (approximately 75\% hydrogen, 25\% helium, and a scattering of other elements), is one of the great successes of cosmology. 
Cosmological measurements on which nucleosynthesis has bearing include anything that would change the nuclear reaction rate, or the duration of this period of nuclear burning.   For example, some of the cosmological data from the CMB and LSS hint at the possibility of an extra relativistic species, like a neutrino, in the early universe.  The presence of such a particle can't change nucleosynthesis much before violating constraints, which strictly limits the type of particle that such an extra relativistic specie could be.  

I could also talk about the wider use of {\bf peculiar velocities}, which are becoming a much more powerful probe, now that  wide area distance surveys of the nearby universe are in progress or coming online soon.  These use techniques such as Tully-Fisher (related to the speed of a galaxy's rotation) and the Fundamental Plane (related to the apparent size and velocity dispersion of elliptical galaxies) to measure distances.  


In the future we may be able to do {\em real-time cosmology} \citep{quercellini12}, by watching the {\bf redshift of a galaxy change over time} \citep{sandage62,liske08} or detecting {\bf cosmological parallax} because our view of galaxy positions shifts as the earth moves with respect to the CMB \citep{kardashev86,ding09}.  
And of course, one of the most exciting windows we're eager to open is the view into the gravitational realm through detection of {\bf  gravitational waves}, indications of which were recently reported by \cite{bicep2_2014}.

\section{Conclusions}

The purpose of this review was to give a general qualitative review of the many varied ways in which dark energy has been measured, and thus the many ways in which theories of dark energy have to be tested before they can be validated.  I find it astonishing, the range and variety of measurements that the standard ${\rm \Lambda}$CDM model has now passed.  One of the difficulties in testing new theories of gravity, or models of dark energy, is that it is not always obvious what effect they will have on all the observational effects above.  Strong collaboration between theorists and the teams of observers making these measurements will be crucial for theory to be robustly tested and to make progress on identifying the nature of dark energy.

\begin{acknowledgements}
I'd like to thank the relatively unsung heroes of this field, the observers and instrument designers, who create the technology that enables us to have this wealth of observational data to challenge and inform our understanding of the universe.  
Many thanks also to those who provided comments on drafts, including Chris Blake, Jason Dossett, Dragan Huterer,  Catherine Heymans, Anthea King, Eric Linder, Jochen Liske, and David Parkinson.  TMD acknowledges the support of the Australian Research Council (ARC) through a Future Fellowship award, FT100100595 and as part of the ARC Centre of Excellence for All-sky Astrophysics (CAASTRO), project  CE110001020.
\end{acknowledgements}



\end{document}